\documentclass[11pt]{article}

\columnwidth\textwidth
\usepackage{graphicx,epsfig}

\begin{document}
\title{Improved effective momentum approximation\\
for\\
quasielastic (e,e') scattering off highly charged nuclei}
\author{Andreas Aste and J\"urg Jourdan\\ 
Institute for Physics, University of Basel, 4056 Basel, Switzerland}
\date{April 29, 2004}

\maketitle

\begin{center}
\begin{abstract}
We address the problem of including Coulomb distortion
effects in inclusive quasielastic $(e,e')$ reactions
using an improved version of the effective momentum approximation.
Arguments are given that a simple modification of
the effective momentum approximation, which is no longer
applicable in its original form in the case of highly charged nuclei,
leads to very good results.
A comparison of the improved effective momentum method with
exact calculations is given. A critical remark concerns
recent experiments with inclusive quasielastic positron
scattering. 
\vskip 0.2 cm
\noindent {\bf{PACS:}} numbers:\\
\noindent 25.70.Bc: Elastic and quasielastic scattering\\
\noindent 25.30.Fj: Inelastic electron scattering to continuum\\
\noindent 11.80.Fv: Approximations (eikonal approximation)
\end{abstract}
\end{center}
\newpage

Nucleon knockout by quasielastic
electron scattering provides a powerful possibility to explore
the electromagnetic properties of nucleons and of the momentum distributions in nuclei, since the transparency of the nucleus with respect to
the electromagnetic probe makes it possible to study the entire nuclear volume. Inclusive scattering provides information on a number of
interesting nuclear properties. The width of the quasielastic peak
allows a measurement of the nuclear Fermi momentum
\cite{Whitney74}, whereas the tail of the quasielastic peak at
low energy loss and large momentum transfer gives information on
high-momentum components in nuclear wave functions \cite{Benhar94a}.
The integral strength of quasielastic scattering,
when compared to sum rules, gives information about the reaction mechanism
and eventual modifications of nucleon form factors in the nuclear medium
\cite{Jourdan96a}. Finally, the scaling properties of the quasielastic
response allows to study the reaction mechanism \cite{Day90},
and extrapolation of the quasielastic response to $A=\infty$
provides us with a very valuable observable of infinite nuclear matter
\cite{Day89}.

The differential cross section for the knockout process can be
written in Born approximation as (for details see \cite{Udias})
\begin{equation}
\frac{d^4 \sigma}{d \epsilon_f d \Omega_f dE_f d\Omega_f} =
\frac{4 \alpha^2 \epsilon_f^2 E_f P_f}{(2 \pi)^5}
\, \delta(\epsilon_i+E_A-\epsilon_f-E_f-E_{A-1})
\bar{\Sigma} |W_{if}|^2, \label{cross}
\end{equation}
where $\bar{\Sigma}$ indicates the sum (average) over final
(initial) polarizations and
\begin{equation}
W_{if}=\int d^3 x \int d^3 y \int \frac{d^3
q}{(2 \pi)^2}
j_\mu^e(\vec{x}) \frac{e^{-i \vec{q} (\vec{x}-\vec{y})}}{q_\mu^2}
J_N^\mu(\vec{y}).
\end{equation}
In this expression, $j_\mu^e$ and $J_N^\mu$ stand for the electron
and nuclear currents, respectively. $\epsilon_{i,f}$
$(E_{i,f})$ are the initial
and final energy of the electron (nucleon), and $P_f$ is the
final momentum of the nucleon. The $\delta$-distribution in 
(\ref{cross}) assures energy conservation for the involved particles
and the (residual) nucleus.
The electron current is given by the well-known Dirac particle expression
\begin{equation}
j_e^\mu(\vec{x})=\bar{\Psi}_e^f (\vec{x}) \gamma^\mu \Psi_e^i (\vec{x}).
\end{equation}
For light nuclei, a description of electron wave functions
by plane waves is a sufficient approximation for many applications,
but for heavy nuclei Coulomb corrections (CC) may become large
and affect the measured cross sections; this needs to be accounted for,
if one aims at a quantitative interpretation of data.
Unfortunately, full distorted wave Born approximation calculations
involving Dirac wave functions lead to an extensive calculational effort.
Kim {\emph{et al.}} proposed a local effective momentum
approximation (LEMA), which leads to good results for heavy nuclei,
but it still necessitates the introduction of non-planar
wave functions \cite{Kim}.

The standard method in the case of light nuclei to handle CC
for elastic scattering in the data
analysis is the effective momentum approximation (EMA).
EMA accounts for two effects of the charged nucleus
on the electron wave function.
Firstly, the initial and final electron momentum $\vec{k}_{i,f}$
is enhanced in the vicinity of
the nucleus due to the attractive electrostatic potential.
Secondly, the attractive potential of the nucleus leads to a focusing of the electron wave function.
For a highly relativistic electron with zero impact parameter
the effective momenta $k_{i,f}'$ of the electron in the center
of the nucleus are given by
\begin{equation}
k_i'=k_i+\Delta k, \quad k_f'=k_f+\Delta k, \quad k_{i,f}=|\vec{k}_{i,f}|,
\quad \Delta k = -V_0/c,
\end{equation}
where $V_0$ is the potential energy of the electron in the center
of the nucleus. The initial and final energy of the electron can
be set equal to $\epsilon_{i,f}=k_{i,f}/c$.
Kim {\emph{et al.}} \cite{Kim} calculated $V_0$ from
the approximate formula
\begin{equation}
V_0=-\frac{3 \alpha Z}{2 r_c}, \quad r_c=[1.1 \, A^{1/3}
+0.86 \, A^{-1/3}] \, \mbox{fm},
\end{equation} 
which is valid for heavy nuclei with charge $Z$ and mass number $A$.
E.g., for $^{208}$Pb we have $V_0=-26.6\, \mbox{MeV}$,
not a negligible quantity when compared to energies of
some hundreds of MeV typically used in electron scattering experiments.

Knoll \cite{Knoll} derived the enhancement factor $F_{i,f}(\vec{r})$
of the electron wave amplitude in the vicinity of the nucleus
from a high energy
partial wave expansion, following previous results given
by Lenz and Rosenfelder \cite{Lenz,Rosenfelder}.
The focusing factor in the center of the nucleus is given
approximately by
\begin{equation}
F_{i,f}(0)=k'_{i,f}/k_{i,f}=\Bigl(1-\frac{V_0}{\epsilon_{i,f}} \Bigr).
\end{equation}

Therefore, EMA corrected cross sections are obtained
by first calculating the cross sections
using plane electron waves but with the electron momenta replaced
by the corresponding effective values.
The result obtained this way must be
multiplied by $F_i^2$, since the focusing of the incoming wave
enters quadratically into the cross section. The focusing factor
for the outgoing wave is automatically generated by the enhanced phase
space factor $\sim {k'_f}^2$ in the effective cross section.
Note that there is also an alternative but equivalent formulation
of EMA \cite{Rosenfelder2}. There, the focusing factors
are automatically absorbed in the Mott cross section which is
used as a prefactor in the full expression for the
knockout cross section. However, the matrix element for the
knockout process is then defined without a photon propagator
$\sim q_\mu^{-2}$.

Kim et al. \cite{Kim}
performed exact calculations for quasielastic electron
scattering using Dirac wave functions both for electrons and nucleons.
The calculations clearly show that EMA has a tendency
to underestimate the cross
sections in relevant kinematical regions which were explored
experimentally at Saclay \cite{Zghiche}. Fig. 1 shows an example
for initial electron energy $\epsilon_i=485$ MeV, scattering
angle $\Theta_e=60^o$ and varying electron energy loss $\omega$.

But a simple modification of the effective momentum approximation,
called EMA' for short in this paper, improves strongly the situation.
The key observation is the fact that most of the nucleons are located
near the surface of the nucleus due to simple geometrical reasons,
where the potential energy of the
electron is given approximately by $2 V_0/3$. But the focusing
of the electron wave is described better by using the central potential 
value $V_0$. The reason for this is the fact that the focusing in
the forefront of the nuclear center (with respect to the direction
of the electron momentum) is lower than in the
center of the nucleus, whereas it is higher in the backfront.
Furthermore, the focusing does not fall off very strongly in
transverse direction \cite{Giusti}.
Therefore, a focusing factor
\begin{equation}
F_{i,f}^2=\Bigl( 1-\frac{V_0}{\epsilon_{i,f}} \Bigr)^2
\end{equation}
is a good average value for the entire nuclear volume,
but for the calculation of the effective cross section,
\begin{equation}
k_i'=k_i+\Delta k', \quad k_f'=k_f+\Delta k',
\quad \Delta k'=-\frac{2}{3c} V_0
\end{equation}
is a better choice for the effective momenta.
EMA' can therefore be expressed by
the following simple recipe: Use $\Delta k= -V_0/c$ for
the calculation of the focusing factors
$F_{i,f}=(k_{i,f}+\Delta k)/k_{i,f}$,
and
$\Delta k'= -2V_0/3c$ for the plane wave calculation
of the cross section. Special attention must be paid
to the fact that the focusing of the outgoing electron
wave function is no longer included properly in the phase space factor
and must therefore be corrected subsequently.

We checked our assumption using the results of Kim {\emph{et al.}}
\cite{Kim}. 
The EMA results displayed in Fig. 1 were obtained from
effective cross sections with a too large $\Delta k=26.6$ MeV/c
instead of using a $\Delta k'=(2/3) \cdot 26.6$ MeV/c.
Our strategy was therefore to modify first
the EMA values by correcting the focusing factor according
to EMA' values $\Delta k'=26.6$ MeV/c and
$\Delta k=(3/2)\cdot 26.6$ MeV/c.
Since the new values obtained this way correspond to a central potential
value which is to large by 50 percent, we shifted then the
modified EMA curve towards the plane wave curve,
such that we obtained
a curve which gives a good estimate for the EMA' values
which would have been obtained by Kim {\emph{et al.}} if they
had applied the EMA' method.
The interpolation procedure used is only approximate,
i.e. we simply moved the modified EMA curve along the
connecting line of the two peaks of the curves by one
third of the distance between the two peaks.
But since the Coulomb distortion is a correction and not
excessively large, higher order effects do not play a crucial role.
The astonishing result of the procedure
is shown in Fig. 2. EMA' and exact values
are in excellent agreement now. We also applied EMA' to the
values given in \cite{Kim} for $\epsilon_i=310$ MeV and
electron scattering angle $\Theta_e=143^o$, with an even
better result in the region with low energy transfer $\omega$.

\begin{figure}[htb]
        \centering
        \includegraphics[width=9cm]{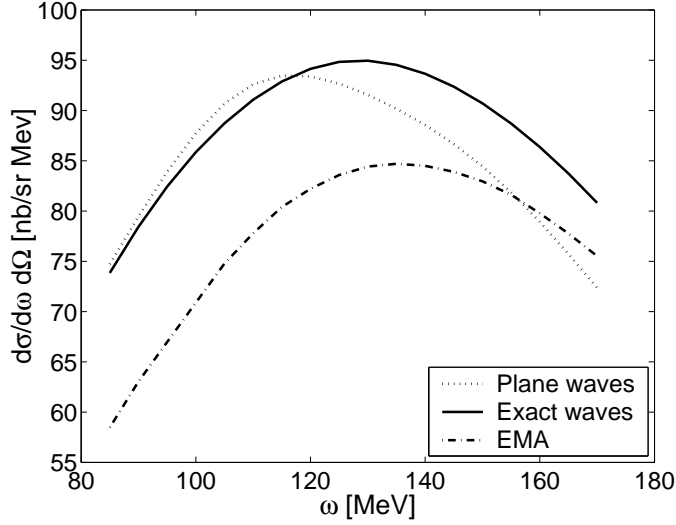}
        \caption{Cross sections for inclusive (e,e') scattering
        on lead in different approaches (taken from \cite{Kim}).
        Cross sections obtained
        from EMA deviate strongly from the results obtained
        by using exact Dirac wave functions for electrons.}
        \label{figcharge}
\end{figure}

\begin{figure}[htb]
        \centering
        \includegraphics[width=9cm]{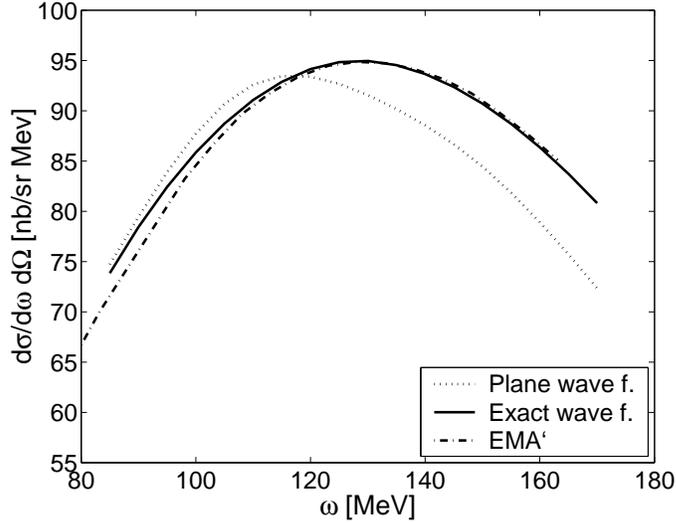}
        \caption{Modified effective momentum approximation:
        EMA' values are in excellent agreement with exact values.}
        \label{figcharge2}
\end{figure}

Gu\`eye {\em{et al.}} \cite{Gueye}
measured positron quasielastic cross sections
on $^{12}$C and $^{208}$Pb and extracted the corresponding
total response function which is defined in plane wave
Born approximation by
\begin{equation}
\frac{d^2 \sigma_{_{PWBA}}}{ d \Omega_f d\epsilon_{f}}=
\sigma_{Mott} \times S^{total}(|\vec{q}|,\omega,\Theta),
\end{equation}
where 
\begin{equation}
\sigma_{Mott}=4 \alpha^2 \cos^2(\Theta/2) \epsilon_f^2/q_\mu^4.
\end{equation}
As mentioned above, the Mott cross section remains unchanged
when it gets multiplied by the EMA focusing factors
and the momentum transfer $q_\mu^4$ is replaced by its
corresponding effective value.
Assuming that EMA is correct, it was found
that the total response functions for electrons with an initial energy
of $383$ MeV and scattering angle
$60^o$ and positrons with an initial energy of
$420$ MeV and the same scattering angle
are nearly identical (see Fig. 3).
This finding is based on the assumption that the relevant
effective momenta are best described by a momentum shift of
approximately $[(420-383)/2]$ MeV $\sim$ $18.5$ MeV both
for electrons and positrons, but with opposite sign.
Whereas the electrons are accelerated to an effective momentum
of approximately $401.5$ MeV/c in the relevant nuclear volume,
the positron momentum is reduced to approximately the same value.

\begin{figure}[htb]
        \centering
        \includegraphics[width=9cm]{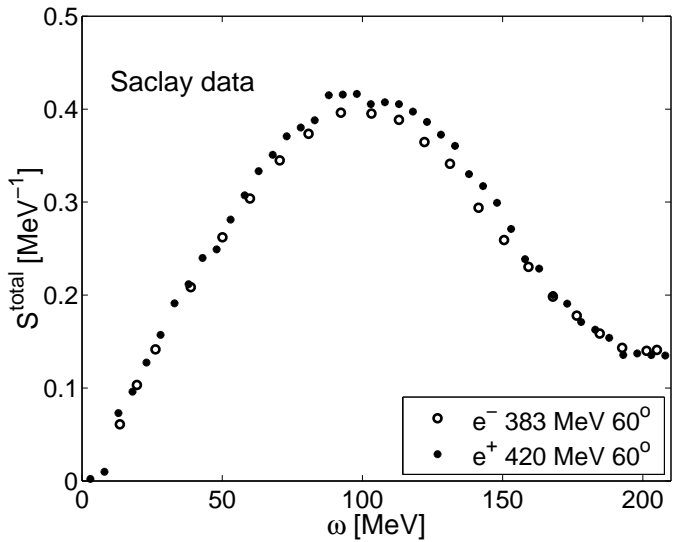}
        \caption{Experimental electron and positron response
         functions given by Gu\`eye {\em{et al.}}.}
        \label{figsaclay}
\end{figure}

In order to substantiate our observation concerning EMA', we
performed calculations using an eikonal approximation for
electron and positron wave functions according to \cite{Aste}.
Since we used a single particle shell model for the description
of the $^{208}$Pb nucleus, our model calculation does not contain
contributions arising from correlations, meson-exchange currents
or inelastic scattering from the nucleons in the nuclear ground
state, but this does not affect the clear result which originates
from the distortion of electron and positron wave functions due
to the Coulomb field of the nucleus.

In our calculations we
have chosen a slightly different initial energy of $385$ MeV for electrons
and $420$ MeV for positrons, corresponding to a momentum
shift from $385$ MeV/c to $402.5$ MeV/c for electrons
and $420$ MeV/c to $402.5$ MeV/c for positrons.
Using a simple EMA analysis, the response
functions for $e^+$ and $e^-$
obtained from the theoretical cross sections differ
by a considerable amount, as shown in Fig. 4.
But reducing the response function for electrons
by a correction factor which accounts for the
enhanced focusing of the electron wave function according
to an EMA' central potential value of $-26$ MeV
and enhancing the response function for positrons
according to a central potential value of $26$ MeV
leads to a satisfactory match of the two response
functions, as shown in Fig. 5.

A reanalysis of the data obtained by Gu\`eye {\em{et al.}}
related to the Coulomb sum rule for quasielastic scattering
\cite{Jourdan2} also indicates that the peak values of the total
response functions shown in Fig. 3 could differ by about 15 per cent.
We suggest therefore that the data analysis of the positron
experiment, which used reference data from an earlier
electron scattering experiment \cite{Zghiche}, should be
revised. One critical point in the analysis is the fact that the
positron beam emittance was six times larger that the emittance
of the direct electron beam, resulting in an uncertainty in the absolute
normalization of the measurement. The data of Zghiche which have been
used to correct for this difference of 5 to 10 per cent, show a peculiar
$\vec{q}$-dependence when the total response function is plotted.
Additionally, the incident energies of $420$ MeV for positrons
and $383$ MeV for electrons which lead to a similar effective momentum
have been obtained from an interpolation procedure which used the
same data.
One has to conclude that the normalization of the data is uncertain
by 15 to 20 per cent.

\begin{figure}[htb]
        \centering
        \includegraphics[width=9cm]{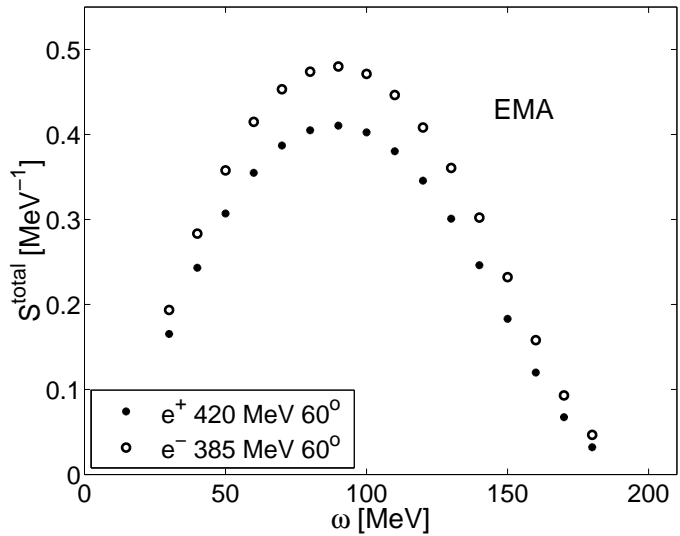}
        \caption{Theoretical response functions for electrons
         and positrons obtained from an EMA analysis.
         The initial energy difference corresponds to a momentum
         shift $385$ MeV/c $\rightarrow$ $402.5$ MeV/c for electrons
         and $420$ MeV/c $\rightarrow$ $402.5$ MeV/c for positrons.}
        \label{figtheo1}
\end{figure}

\begin{figure}[htb]
        \centering
        \includegraphics[width=9cm]{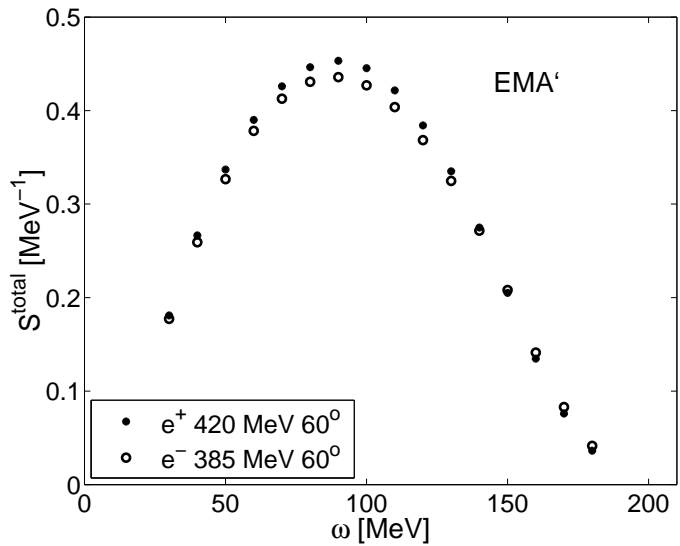}
        \caption{Theoretical response functions for electrons
         and positrons, but EMA' corrected.}
        \label{figtheo2}
\end{figure}

We finally remark that the choice
$\Delta k=-V_0/c$ and
$\Delta k'=-2 V_0/3c$ is an {\emph{ad hoc}} prescription,
but it is motivated from the physical picture that
most of the nucleons of heavy nuclei are located near
the surface of the nucleus.
It clearly remains desirable to have access to exact calculations,
but our calculations using the eikonal approximation for electron
and positron wave functions
and the exact calculations by Kim {\emph{et al.}} 
clearly suggest that EMA is unreliable in the case of heavy nuclei,
whereas EMA' is a reliable approximation for the involved treatment
of CC.

\section*{Acknowledgements}
We thank Kyungsik Kim and Yanhe Jin for providing us with
numerical values of their calculations. This work has been
supported by the Swiss National Science Foundation.

\end{document}